\documentclass[fleqn,twoside]{article}
\usepackage{amsmath}
\usepackage{espcrc2,psfig,epsfig}
\usepackage{graphicx}
\usepackage[figuresright]{rotating}
\newcommand{\D}{\displaystyle}

\newcommand{\minerva}{MINER$\nu$A}

\newcommand{\integ}{\iint}

\newcommand{\fa}{$F_A(q^2)$}
\newcommand{\gep}{$G_E^p(q^2)$}

\title{Vector and Axial Form Factors Applied to Neutrino Quasielastic 
       Scattering}

\author{H. Budd\address[roc]{Department of Physics and Astronomy,
             University of Rochester,
             Rochester, New York 14618,  USA}, A. Bodek\addressmark[roc]
        and
        J. Arrington\address[arg]{Argonne National Laboratory, Argonne,
 Illinois 60439, USA}}
    
\begin{document}

\begin{abstract}
We calculate the 
quasielastic cross sections for neutrino scattering on nucleons using
up to date fits to the nucleon elastic electromagnetic form factors 
$G_E^p$, $G_E^n$, $G_M^p$, $G_M^n$,
and weak form factors. We show the extraction of \fa\ for neutrino
experiments. We show how well \minerva, a new approved 
experiment at FNAL, can measure \fa. We show the that \fa\ has a different
contribution to the anti-neutrino cross section, and 
how the anti-neutrino data can be used to check 
\fa\ extracted from neutrino scattering.
(Presented by Howard Budd at NuInt04,   Mar. 2004, Laboratori Nazionali del Gran Sasso -
INFN - Assergi, Italy~\cite{nuint04})
\vspace{1pc}
\end{abstract}

\maketitle

\section{INTRODUCTION}
Experimental  evidence for oscillations among the three
neutrino generations has been recently reported~\cite{Fukada_98}.
Since quasielastic (QE) scattering forms an important component of 
neutrino scattering
at low energies, we have undertaken to investigate 
QE neutrino scattering using the latest information
on nucleon form factors. 

Recent experiments at SLAC and Jefferson Lab (JLab) have 
given precise measurements of the vector electromagnetic
form factors for the proton and neutron. 
These form factors can be related to the form factors for QE
neutrino scattering by conserved vector current hypothesis, CVC. 
These more recent form factors can be used to give better predictions
for QE neutrino scattering and better determination of the 
axial form factor, \fa.

\section{EQUATIONS FOR QE SCATTERING}
The hadronic current for QE neutrino scattering is given by~\cite{Lle_72}
\begin{eqnarray*}
 \lefteqn{<p(p_2)|J_{\lambda}^+|n(p_1)>  =   } \nonumber \\
& \overline{u}(p_2)\left[
  \gamma_{\lambda}F_V^1(q^2)
  +\frac{\D i\sigma_{\lambda\nu}q^{\nu}{\xi}F_V^2(q^2)}{\D 2M} \right. \nonumber \\
& \left. ~~~~~~~~~~~+\gamma_{\lambda}\gamma_5F_A(q^2)
+\frac{\D q_{\lambda}\gamma_5F_P(q^2)}{\D M} \right]u(p_1),
\end{eqnarray*}
where $q=k_{\nu}-k_{\mu}$, $\xi=(\mu_p-1)-\mu_n$, and 
$M=(m_p+m_n)/2$.  Here, $\mu_p$ and $\mu_n$ are the 
proton and neutron magnetic moments.
We assume that there are no second class currents, so the scalar
form factor  $F_V^3$ and the tensor form factor $F_A^3$
need not be included. 

\begin{figure*}[htb]
\begin{center}
\epsfxsize=6.51in
\mbox{\epsffile[0 40 567 260]{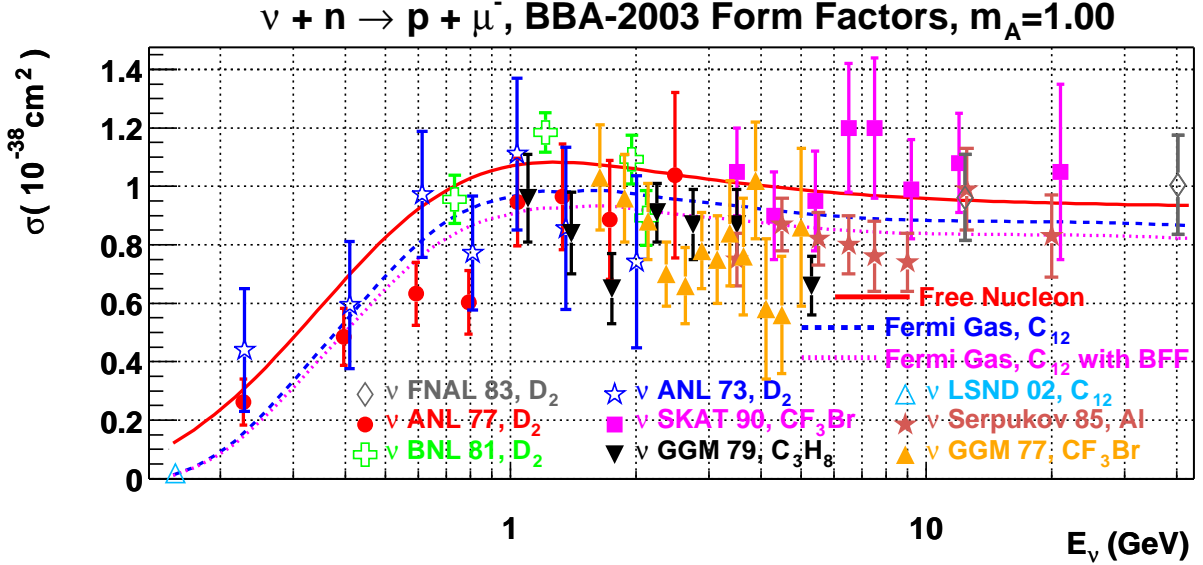}}
\end{center}
\caption{The QE neutrino cross section along 
with data from various experiments. The calculation uses 
$M_A$=1.00 GeV, $g_A$=$-1.267$, $M_V^2$=0.71 GeV$^2$ and 
BBA-2003 Form Factors.
The solid curve uses no nuclear correction,
while the dashed curve~\cite{Zeller_03} uses
a Fermi gas model for carbon with a 25 MeV binding energy
and 220 Fermi momentum. The dotted curve is the prediction for carbon
including both Fermi gas Pauli blocking and the effect of
nuclear binding on the nucleon form factors~\cite{Tsushima_03}(bounded
form factors).
The data shown are from 
FNAL 1983~\cite{Kitagaki_83},
ANL 1977~\cite{Barish_77},
BNL 1981~\cite{Baker_81},
ANL 1973~\cite{Mann_73},
SKAT 1990~\cite{Brunner_90},
GGM 1979~\cite{Pohl_79},
LSND 2002~\cite{Auerbach_02},
Serpukov 1985~\cite{Belikov_85},
and GGM 1977~\cite{Bonetti_77}.}
\label{elas_JhaKJhaJ_nu}
\end{figure*}

The form factors $ F^1_V(q^2)$ and  ${\xi}F^2_V(q^2)$
are given by:
$$ F^1_V(q^2)=
\frac{G_E^V(q^2)-\frac{\D q^2}{\D 4M^2}G_M^V(q^2)}{1-\frac{\D q^2}{\D 4M^2}},
$$
$$
{\xi}F^2_V(q^2) =\frac{G_M^V(q^2)-G_E^V(q^2)}{1-\frac{\D q^2}{\D 4M^2}}.
$$

We use the CVC to determine $ G_E^V(q^2)$ and $ G_M^V(q^2)$ 
from  the electron scattering form factors
$G_E^p(q^2)$, $G_E^n(q^2)$, $G_M^p(q^2)$, and $G_M^n(q^2)$:

$$ 
G_E^V(q^2)=G_E^p(q^2)-G_E^n(q^2), 
$$
$$
G_M^V(q^2)=G_M^p(q^2)-G_M^n(q^2). 
$$

Previously, many neutrino experiment have assumed
that the vector 
form factors are 
described by the dipole approximation.
$$ G_D(q^2)=\frac{1}{\left(1-\frac{\D q^2}{\D M_V^2}\right)^2 },~~M_V^2=0.71~GeV^2$$
$$ 
G_E^p=G_D(q^2),~~~G_E^n=0,
$$ 
$$ 
G_M^p={\mu_p}G_D(q^2),~~~ G_M^n={\mu_n}G_D(q^2).
$$

We refer to the above combination of form factors
as `Dipole Form Factors'. It is an approximation that
has been improved by us in a previous publication~\cite{Budd:2003wb}.
We use our updated form factors  which we refer
 as `BBA-2003 Form Factors'~\cite{Budd:2003wb}~\cite{JRA_03}
 (Budd, Bodek, Arrington).

The axial form factor is given by 
$$ F_A(q^2)=\frac{g_A}{\left(1-\frac{\D q^2}{\D M_A^2}\right)^2 }. $$
We have used our updated  value of 
$M_A$=1.00 $\pm$ 0.020 GeV~\cite{Budd:2003wb}  which is in good agreement with the
theoretically corrected value
from pion electroproduction of 1.014 $\pm$ 0.016 GeV~\cite{Bernard_01}.
For extraction of \fa\ we use the value of $M_A$ = 1.014, since it is
independent of QE scattering measurements.

\section{Comparison to Cross Section Data}
Figures~\ref{elas_JhaKJhaJ_nu}
shows the QE cross section for 
$\nu$ using BBA-2003 Form Factors and $M_A$=1.00 GeV
The normalization uncertainty in the data is approximately 10\%.
The solid curve uses no nuclear correction, 
while the dotted curve~\cite{Zeller_03} uses
a NUANCE~\cite{Casper_02} calculation of a 
Smith and Moniz~\cite{Smith_72} based Fermi gas model
for carbon. This nuclear model includes Pauli blocking and Fermi
motion, but not final state interactions. 
The Fermi gas model was run with a 25 MeV binding energy and
220 MeV Fermi momentum.
The dotted curve is the prediction for carbon including
both Fermi gas Pauli blocking 
and the effect of nuclear binding on the nucleon form factors
as modeled by Tsushima {\em et al.}~\cite{Tsushima_03}.
The ratio of bounded form factors to free form factors is
set to 1 for $Q^2>2.0~GeV$.
The updated form factors improve the agreement 
with neutrino QE cross section data 
and give a reasonable description of the 
cross sections from deuterium.  
We plan to study the nuclear corrections, adopting models
which have been used in precision electron scattering measurements
from nuclei at SLAC and JLab.

\begin{figure}[t]
\vspace{9pt}
\psfig{figure=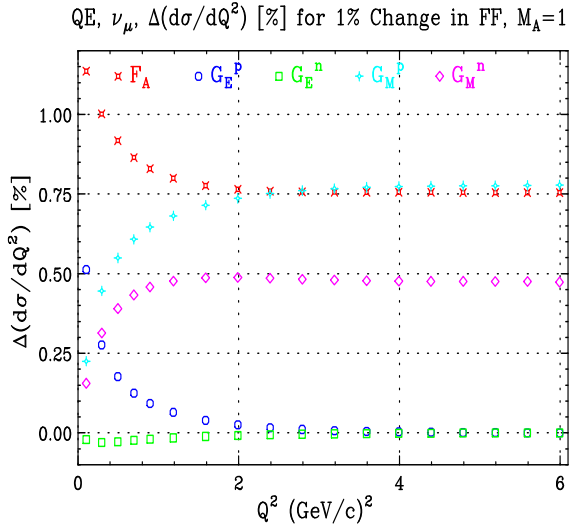,width=2.94in,height=2.74in}
\vspace{-1.11cm}
\caption{ The percent change in the neutrino cross section for
a 1\% change in the form factors. }
\label{ddsigma_dq_dff_nu}
\end{figure}

\begin{figure}[t]
\vspace{9pt}
\psfig{figure=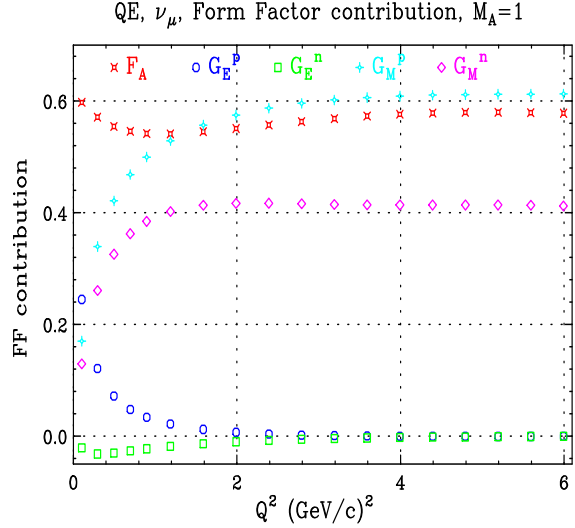,width=2.94in,height=2.74in}
\vspace{-1.11cm}
\caption{ Fractional contribution of the form factor  determined by setting the
form factor to zero and by determining the fractional 
decrease in the differential cross section, 
$ 1 -  (d\sigma/dQ^2(form factor=0))/(d\sigma/dQ^2)$.}
\label{FF_contribution_nu}
\end{figure}

\vspace{-0.2cm}

\section{Extraction of $F_A(q^2)$}
A  substantial fraction 
of the cross section comes from the form factor \fa.
Therefore, we can extract \fa\ from the differential
cross section.
Figure~\ref{ddsigma_dq_dff_nu}~and~\ref{FF_contribution_nu} 
show the contribution of \fa\ 
to $d\sigma/dQ^2$. Figure~\ref{ddsigma_dq_dff_nu}
shows the percent change in the neutrino cross section for
a 1\% change in the form factors.
  Figure~\ref{FF_contribution_nu} shows the
fractional contribution of the form factor determined by setting the
form factor to zero and by determining the 
fractional decrease in the differential
cross section.  Since some terms are products of different form factors, the sum of 
the curves do not have be 1.  

To extract $F_A$, we write the equation
for $d\sigma/dq^2(q^2,E_\nu)$ in terms of a quadratic 
function of $F_A(q^2)$. 
\begin{eqnarray*}
a(q^2,E_\nu)F_A(q^2)^2+b(q^2,E_\nu)F_A(q^2)~~~~~~~~~~~~~~~~~~~ \\
~~~~~~~~~~~~~~+c(q^2,E_\nu)-\frac{d\sigma}{dq^2}(q^2,E_\nu)=0~~
\end{eqnarray*}
For each $q^2$ bin, we integrate the above equation
over the $q^2$ bin and the neutrino flux.
\begin{eqnarray*}
\integ dq^2dE_\nu \{a(q^2,E_\nu)F_A(q^2)^2+b(q^2,E_\nu)F_A(q^2)\\
+c(q^2,E_\nu)-\frac{d\sigma}{dq^2}(q^2,E_\nu)\}=0
\end{eqnarray*}

\begin{figure}[t]
\vspace{9pt}
\psfig{figure=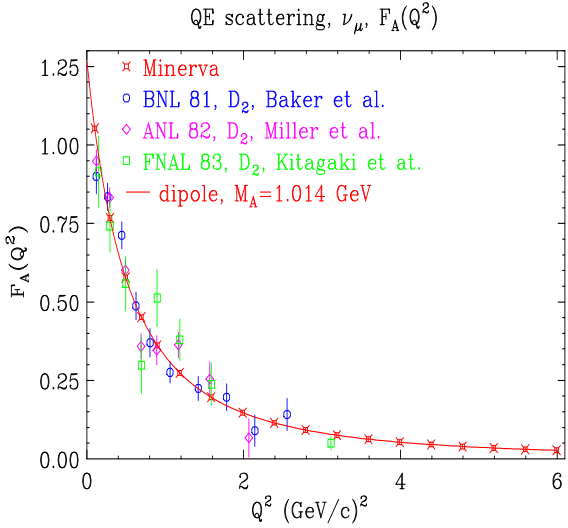,width=2.94in,height=2.74in}
\vspace{-1.11cm}
\caption{ Extracted values of $F_A(q^2)$ for the 
three  deuterium bubble chamber experiments
Baker {\em et al.}~\cite{Baker_81},
Miller {\em et al.}~\cite{Miller_82},
and Kitagaki {\em et al.}~\cite{Kitagaki_83}.
Also shown are the  expected errors for \minerva\
assuming a dipole form factor for \fa\ with M$_A$=1.014.}
\label{f_a_lin_mva}
\end{figure}

The above equation can be written as a quadratic equation 
in $F_A$ at the bin value  $q_{bin}^2$. 
\begin{equation}
\alpha F_A(q_{bin}^2)^2+\beta F_A(q_{bin}^2)+\gamma-\Delta-N_{Bin}^{Data}=0
\nonumber
\end{equation}
  The terms of this equation are given below:
\begin{equation}
\alpha= \integ dq^2dE_\nu a(q^2,E_\nu)
\nonumber
\end{equation}
\begin{equation}
\beta= \integ dq^2dE_\nu b(q^2,E_\nu)
\nonumber
\end{equation}
\begin{equation}
\gamma= \integ dq^2dE_\nu c(q^2,E_\nu)
\nonumber
\end{equation}
To find $q_{bin}^2$, we assume a nominal $F_A(q^2)$, written
$F_A^N(q^2)$. We determine  $q_{bin}^2$ from 
\begin{equation}
\alpha F_A^N(q_{bin}^2)^2-\integ dq^2dE_\nu a(q^2,E_\nu)F_A^N(q^2)^2=0.
\nonumber
\end{equation}
$\Delta$ is a bin center correction term which 
also uses $F_A^N(q^2)$.
$\Delta$ is determined by
\begin{equation}
\Delta=\beta F_A^N(q_{bin}^2)-\integ dq^2dE_\nu b(q^2,E_\nu)F_A^N(q^2).
\nonumber
\end{equation}
The number of events in the bin is given by $N_{Bin}^{Data}$. 
The number of events in the bin from theory is
\begin{equation}
N_{Bin}^{Thy}= \integ dq^2dE_\nu \frac{d\sigma}{dq^2}(q^2,E_\nu).
\nonumber
\end{equation}
The errors in the points are given by 
\begin{equation}
\frac{\sqrt{N_{Bin}^{Thy}}}{2\alpha F_A^N(q_{bin}^2)+\beta}.
\nonumber
\end{equation}

\begin{figure}[t]
\vspace{9pt}
\psfig{figure=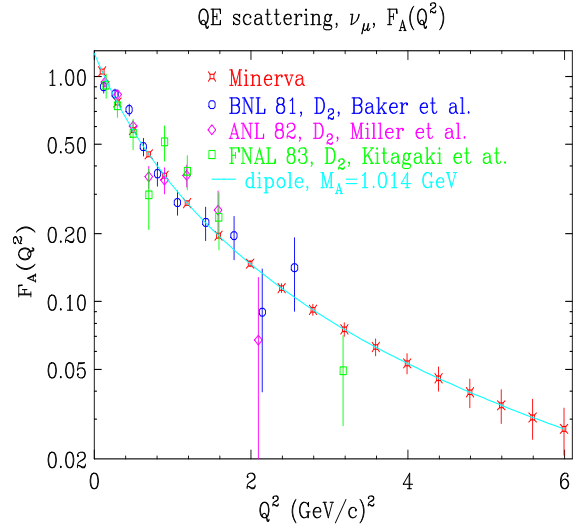,width=2.94in,height=2.74in}
\vspace{-1.11cm}
\caption{ Same as Figure~\ref{f_a_lin_mva} with a logarithmic scale. }
\label{f_a_log_mva}
\end{figure}

Figure~\ref{f_a_lin_mva}~and~\ref{f_a_log_mva} show
our extracted values of \fa\ for the three
deuterium bubble chamber experiments.
For these plots the curve shown in the
figures is a dipole with $m_A$=1.014, the value extracted from 
pion-electro production. The data and fluxes  given in their papers 
are used in the extraction of \fa. 
These plots show the previous data is not sufficient to determine 
the form for \fa. 

In addition, we have shown the expected values for \minerva\
and its errors. We have plotted \minerva\ assuming it is a dipole.
We have assumed a 4 year run with 3 tons of fiducial volume
and included the effects of inefficiencies and 
backgrounds. Resolution smearing and systematic errors are not included.

\begin{figure}[t]
\vspace{9pt}
\psfig{figure=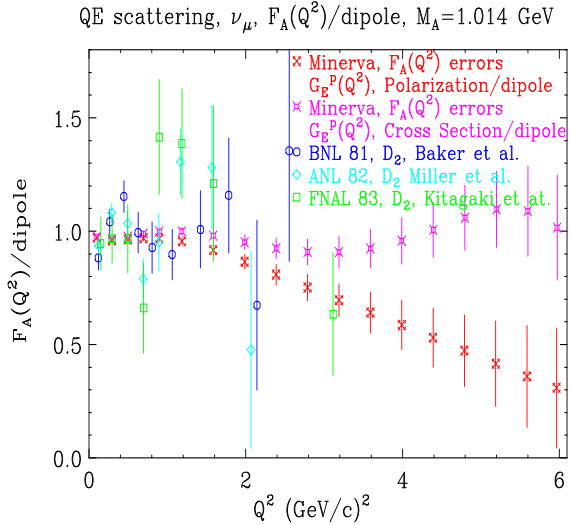,width=2.94in,height=2.74in}
\vspace{-1.11cm}
\caption{Extracted values of $F_A(q^2)/dipole$ for deuterium bubble 
chamber experiments
Baker {\em et al.}~\cite{Baker_81},
Miller {\em et al.}~\cite{Miller_82}, 
and Kitagaki {\em et al.}~\cite{Kitagaki_83}. 
For \minerva\ the projected results are shown for two different 
assumptions: $F_A/$dipole=$G_E^p/$dipole from cross section and 
$F_A/$dipole=$G_E^p/$dipole from polarization. 
The \minerva\ errors are for a 4 year run. }
\label{f_a_mva}
\end{figure}

Figure~\ref{f_a_mva} plots \fa/dipole to show
how well \minerva\ can measure \fa.
\gep\ from electron scattering experiments 
depends upon the measuring technique~\cite{JRA_03}. For 
\minerva\, we show $F_A$ under the assumption  \fa$/$dipole = \gep$/$dipole from 
the cross section technique (Rosenbuth separation) 
and \fa$/$dipole = \gep$/$dipole
from polarization transfer technique. The \minerva\ errors are plotted assuming
the plotted \fa\ is the nominal \fa. We see that the measurement
of \fa\ from \minerva\
can distinguish between these to the two possible forms. 
In addition, \minerva\ can determine whether \fa\ is 
a dipole or not. 

\begin{figure}%[htb]
\vspace{9pt}
\psfig{figure=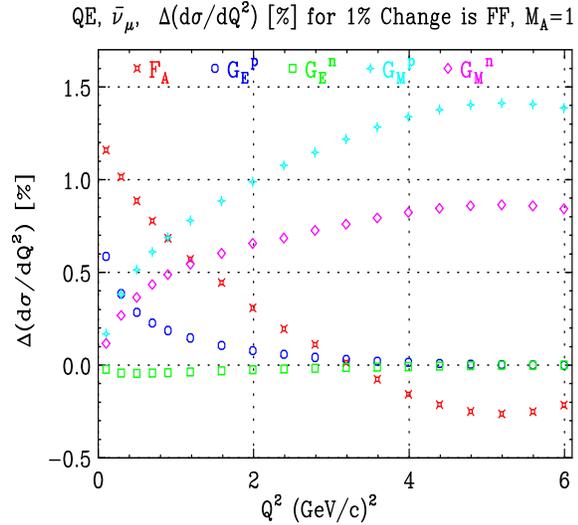,width=2.94in,height=2.74in}
\vspace{-1.11cm}
%\begin{center}
%\epsfxsize=2.94in
%\mbox{\epsffile[51  288  538  571]{ddsigma_dq_dff_nub.eps}}
%\end{center}
\caption{ The percent change in the anti-neutrino cross section for
a 1\% change in the form factors.}
\label{ddsigma_dq_dff_nub}
\end{figure}

\section{Extraction of $F_A(q^2)$ from anti-neutrinos}
\begin{figure}[htb]
\vspace{9pt}
\psfig{figure=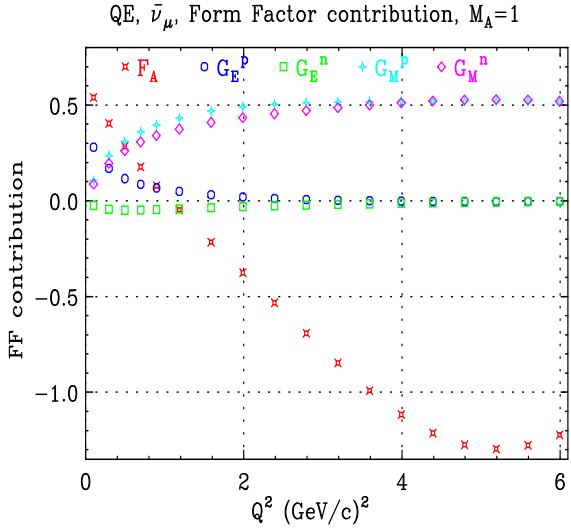,width=2.94in,height=2.74in}
\vspace{-1.11cm}
\caption{$ 1 -  d\sigma/dQ^2(form factor=0)/d\sigma/dQ^2$.
The contribution of the form factors determined 
by setting the form factors = 0.}
\label{FF_contribution_nub}
\end{figure}

   The determination of \fa\ will have  systematic errors from the flux, 
nuclear effects, QE identifications, background determination,
 etc. Anti-neutrino data can provide a check on \fa. 
Figure~\ref{ddsigma_dq_dff_nub}~and~\ref{FF_contribution_nub} 
show the contribution of \fa\
to the cross section vs $Q^2$ for 
anti-neutrinos. Figure~\ref{ddsigma_dq_dff_nub}
shows the percent change in the anti-neutrino cross section for
a 1\% change in the form factors.
The plot shows that \fa\ has a different contribution to 
the cross section for anti-neutrinos than neutrinos.
At  $Q^2 \sim 3 GeV^2$, $F_A$ is not contributing to 
the cross section, and the cross section becomes
independent of \fa. Hence, at higher $Q^2$ the
cross section can be predicted and compared 
to the data to determine errors to the neutrino
extraction.  Figure~\ref{FF_contribution_nub} shows the 
fractional contribution of the form factor determined by setting the
form factor to zero and by determining the 
fractional decrease in the differential cross section. 
Note, since some terms are products of different 
form factors the sum of
the curves do not have to sum to 1.

\begin{figure}[htb]
\vspace{9pt}
\psfig{figure=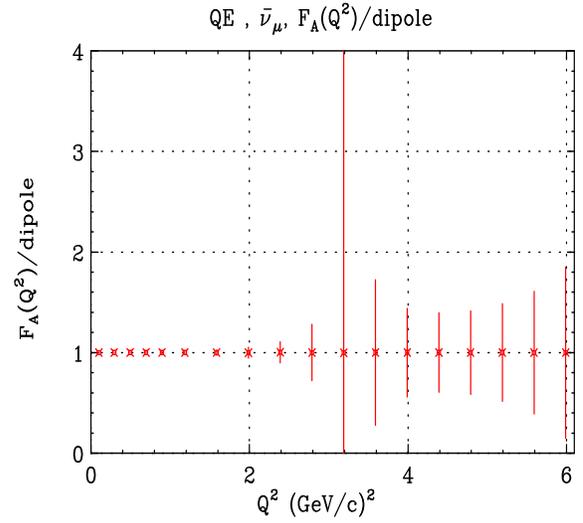,width=2.94in,height=2.7in}
\vspace{-1.11cm}
\caption{The relative errors for an extraction
of $F_A$ using anti-neutrinos. The errors are shown for 
$F_A(Q^2)/dipole$. The flux is arbitrary. }
\label{f_a_dipole_nub}
\end{figure}

Figure~\ref{f_a_dipole_nub} shows the errors on 
$F_A$/dipole for anti-neutrinos. The overall errors scale is arbitrary.
As we expect, the errors on \fa\ become large at $Q^2$ 
around 3 $GeV^2$ when the derivative of the cross 
section with respect to \fa\ goes to 0.

\vspace{-0.1cm}
\section{Conclusions}
We have used new form factors to show 
the cross sections for QE neutrino scattering.
The cross sections give a reasonable description 
of the  deuterium data, but the nuclear data is low.
We have shown how to extract $F_A$ and have shown show well \minerva\
can measure \fa. For anti-neutrino data at high $Q^2$, $F_A$ has 
a different contribution, so anti-neutrinos provides a check for
the extraction of $F_A$ from neutrinos.

\section{Acknowledgments}
This work is supported in part by the U. S. Department of Energy, Nuclear
Physics Division, under contract W-31-109-ENG-38 (Argonne)
and High Energy Physics Division under
grant DE-FG02-91ER40685 (Rochester).

\end{document}